\documentclass[aps,prl,reprint,superscriptaddress]{revtex4-1}

\usepackage{amsmath}    
\usepackage{graphicx}   
\usepackage{color}      
\usepackage{subfigure}  
\usepackage{todonotes}
\usepackage{nicefrac}
\usepackage[utf8]{inputenc}

\begin{document}

\title{Exciton properties of selected aromatic hydrocarbon systems}

\author{Friedrich Roth}
\author{Benjamin Mahns}
\author{Silke Hampel}
\author{Markus Nohr}
\affiliation{IFW Dresden, P.O. Box 270116, D-01171 Dresden, Germany}
\author{Helmuth Berger}
\affiliation{Institute of Physics of Complex Matter, Ecole Polytechnique Federale de Lausanne (EPFL), CH-1015 Lausanne, Switzerland}
\author{Bernd B\"uchner}
\affiliation{IFW Dresden, P.O. Box 270116, D-01171 Dresden, Germany}
\affiliation{Institut f\"ur Festk\"orperphysik, Technische Universit\"at Dresden, D-01062 Dresden, Germany} 
\author{Martin Knupfer}
\affiliation{IFW Dresden, P.O. Box 270116, D-01171 Dresden, Germany}

\date{\today}

\begin{abstract}
We have examined the singlet excitons in two representatives of acene-type (tetracene and pentacene) and phenacene-type (chrysene and picene) molecular crystals, respectively, using electron energy-loss spectroscopy at low temperatures. We show that the excitation spectra of the two hydrocarbon families significantly differ. Moreover, close inspection of the data indicates that there is an increasing importance of charge-transfer excitons at lowest excitation energy with increasing length of the molecules.
\end{abstract}

\maketitle

\section{Introduction}
Pentacene as well as other members of the acene family are famous organic semiconductors. Single crystals of these materials can be grown in a very high quality, and thus very high charge carrier mobilities could be achieved in e.\,g. pentacene and tetracene \cite{Karl2003}. These have been exploited in organic field effect transistors in view of fundamental as well as applied aspects \cite{Takahashi2007,Braga2009,Sirringhaus2009}. In fact, such materials are very promising for a wide range of applications, for
instance, in (opto)electronic devices such as field-effect transistors \cite{Gershenson2006}, light-emitting diodes \cite{Baldo1998,Friend1999}, and photovoltaic cells \cite{Peumans2003,Li2005}.

\begin{figure}[h]
\centering
\includegraphics[width=0.9\linewidth]{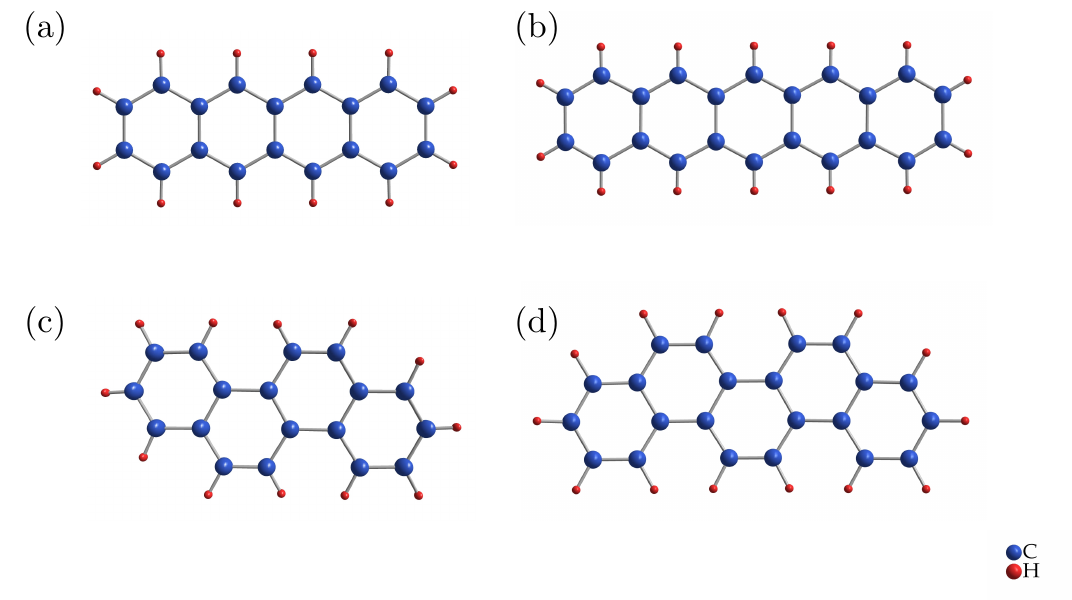}
\caption{Schematic representation of the molecular structure of the investigated hydrocarbons. (a) Tetracene, (b) Pentacene, (c) Chrysene, and (d) Picene} \label{f1}
\end{figure}

The schematic representation of the molecular structure of the aromatic hydrocarbons investigated in this paper is shown in Fig.\,\ref{f1}.
Tetracene and pentacene belong to the so-called acenes and are characterized by a linear arrangement of the benzene rings. In contrast, chrysene and picene are close relatives but the benzene rings are arranged in a zigzag manner as depicted in Fig.\,\ref{f1} lower panel. The latter have received increasing attention recently due to the observation of superconductivity in alkali doped hydrocarbon crystals, among them picene, with relatively high transition temperatures to the superconducting state \cite{Mitsuhashi2010,Wang2011,Wang2011_2,Xue2011,Ying2012,Kasahara2012}. In all cases the molecules in the solid arrange in a herringbone
manner which is typical for many aromatic hydrocarbon molecular solids and can be seen as a consequence of the densest possible packing of the molecules within the crystal with the least possible repulsion.

\par

In this contribution we present an investigation of the electronic excitation spectra of tetracene, pentacene, chrysene and picene single
crystals, which have been measured using electron energy-loss spectroscopy (EELS) in transmission. This technique can probe the excitations as a function of energy and momentum, which allows additional information on the nature of electronic excitations as compared to optical studies \cite{Knupfer2000,Schuster2007,Roth2011,Cudazzo2011,Roth2012}. We demonstrate that the excitation spectra of chrysene and picene on the one hand, and tetracene and pentacene on the other hand differ substantially. Moreover, we discuss the importance of charge-transfer excitations at low energies.

\section{Experimental}

\begin{figure*}
 \centering
\includegraphics[width=0.9\linewidth]{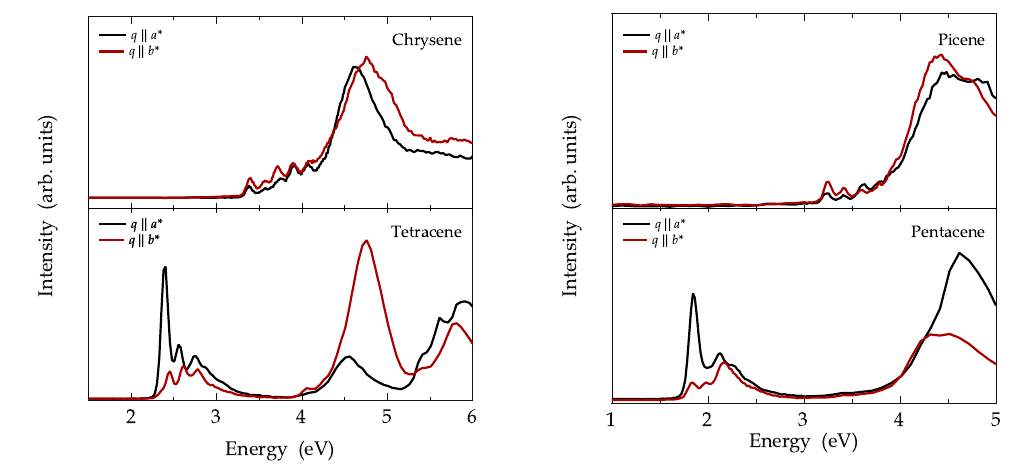}
\caption{Loss function measured along the two fundamental crystallographic axis, $a^*$ and $b^*$, for chrysene and tetracene (left panel) and picene and pentacene (right panel). All measurements have been performed at a small momentum transfer of $q$\,=\,0.1\,\AA$^{-1}$ and at 20 K.}
\label{f2}
\end{figure*}

Pentacene single crystals with a high quality and typical crystal dimensions of 5-15\,mm length, 2-5\,mm width, and 0.05-0.25\,mm thickness were obtained via a directional sublimation of two or three times purified pentacene (Fluka). The crystal growth was carried out at temperatures between 280\,$^{\circ}\mathrm{C}$ and 220\,$^{\circ}\mathrm{C}$ in closed, evacuated pyrex ampoules and a horizontal two-zone furnace was used. The growth procedure lasted about four to six weeks. In case of tetracene (Sigma-Aldrich Chemie GmbH), single crystals were synthesized via physical transport in an inert gas stream (argon and hydrogen mixture), whereas the sublimation occurred at about 300\,$^{\circ}\mathrm{C}$ in the hot zone of the furnace and the tetracene single crystals were grown at a temperature of about 150\,-\,200\,$^{\circ}\mathrm{C}$. To obatin well defined and large crystals (dimension of 10\,mm x 7\,mm x 0.1\,mm) the crystal growth lasted between one and six hours. The chrysene and picene single crystals (both Sigma-Aldrich Chemie GmbH) were prepared via physical vapor growth in a vertical geometry. Chrysene as well as picene were sublimed from a glass surface and the crystal growth occurred on a Al foil on top. The growth lasted 12 hours and resulted in very thin, singly-crystalline platelets with typical dimensions of about 0.5\,mm x 0.5\,mm x 100\,nm.

\par

Our EELS investigations require thin samples (thickness $\sim$ 100\,nm) which in the case of tetracene and pentacene were cut from the flat
surface of a single crystal platelet with an ultramicrotome using a diamond knife. The as-grown picene and chrysene crystals were thin enough and could be used without further preparation. All samples were attached to standard electron microscopy grids \cite{Fink1994}, mounted into a sample holder, and transferred into the EELS spectrometer. All EELS measurements were carried out using a 172\,keV spectrometer described elsewhere \cite{Fink1989}. At this high primary beam energy only singlet excitations are possible \cite{Fink1989}. The energy and momentum resolution were 85\,meV and 0.03\,\AA$^{-1}$, respectively. The EELS signal, which is proportional to the loss function $\operatorname{Im}(-1/\epsilon({q},\omega$)), was determined for momentum transfers, $q$, parallel to the directions of the corresponding reciprocal lattice vectors [$\epsilon(q,\omega)$ is the dielectric function]. Moreover, a He flow cryostat allows to cool the sample down to 20\,K.

\par

Prior to the investigation of the electronic excitations, our films were thoroughly characterized \emph{in-situ} using electron diffraction. These investigations clearly document that the films are single crystalline and of very good quality. Furthermore, as
molecular crystals often are damaged by fast electrons, we repeatedly checked our samples for any sign of degradation. In particular,
degradation was followed by watching an increasing amorphous-like background in the electron diffraction spectra and an increase of spectral
weight in the loss function in the energy region below the first excitation feature. It turned out that under our measurement conditions the
spectra remained unchanged for about 12\,h. Samples that showed any signature of degradation were not considered further but replaced by newly
prepared thin films. The results from the different films have been shown to be reproducible.

\section{Results and discussion}

We start the presentation of our results with a comparison of the loss functions for momentum vectors $q$ parallel to the reciprocal
lattice directions $a^*$ and $b^*$ and a small momentum transfer of 0.1\,\AA$^{-1}$ as shown in Fig.\,\ref{f2}. At such a small momentum
transfer we probe essentially vertical transitions, the so-called optical limit, i.\,e., a comparison to optical data is possible. We note
however that our data are measured along the reciprocal lattice directions $a^*$ and $b^*$, which due to the low crystal symmetry of the acene and phenacene crystals are not parallel to the crystal axes $a$ and $b$. Fig.\,\ref{f2} clearly demonstrates the differences in the electronic excitation spectra of the two types of hydrocarbons. For chrysene and picene we observe a broad peak around 4.5\,eV as well as a pronounced fine structure right after the excitation onset. In addition, the optical gap is about 3.3\,eV for chrysene and 3.2\,eV for picene, respectively. The main features in our spectra are in very good agreement with previous optical absorption data as well as EELS studies of crystals or polycrystalline films \cite{Tanaka1965,Roth2011_exci}. Often, the low energy electronic excitations in molecular solids are excitons, i.\,e. bound electron-hole pairs with a sizeable binding energy \cite{Hill2000,Knupfer2003,Pope1999}. This is also true for the two phenacenes studied here since their excitation onset is significantly below the transport energy
gap which has been estimated to be about 4.2\,eV for chrysene and 4\,eV for picene \cite{Sato1987,Roth2010}. The eqiuvalent is true for tetracene and pentacene with transport gaps of about 3.3\,eV and 2.2\,eV, respectively \cite{Amy2005,Sato1981,Nayak2009}. For many molecular crystals it is not unusual that the lowest singlet excitations are split into into Davydov components, which arise from the interaction of the excitation dipoles of the symmetrically inequivalent molecules in the crystal unit cell \cite{Davydov1971,Pope1999}. These Davydov components frequently have a particular polarization dependence. The spectra for chrysene and picene however as shown in Fig.\,\ref{f2} are isotropic in terms of the energy positions, and only some small intensity variations are seen going from the $a^*$ to the $b^*$ directions. Thus, our data do not reveal different Davydov components. This might be related to the rather small spectral weight of the lowest lying excitations in these materials, which could also imply a rather small Davydov splitting \cite{Davydov1971,Pope1999}.

\par

Going to the linear acene crystals (lower panels in Fig.\,\ref{f2}), these observations change substantially. The low energy excitations occur at much lower energies, their spectral weight is much larger, and their fine structure is significantly different. From systematic studies in the past of the optical absorption spectra of acenes and phenacenes as a function of the molecule lengths it is known that the energy position of these excitations in acenes quite strongly shifts to lower energies upon increasing the molecules, while it remains quite constant in the case of the phen\-acenes \cite{Wiberg1997}. This is related to the molecular structure and symmetry, which allows a delocalization of the highest occupied molecular orbital (HOMO) and the lowest unoccupied molecular orbital (LUMO) of acenes over the entire molecules, while for their counterparts these orbitals remain localized at particular sites of the molecule \cite{Hiruta1997,Wiberg1997}. This difference in the molecular orbital structures most likely is also responsible for the different excitation probabilities (intensities), since this is directly related to the overlap of these orbitals. In view of the different excitation intensities it is also tempting to attribute the clearly observed Davydov splitting for tetracene (with Davydov components at about 2.38\,eV and 2.45\,eV) and pentacene (with Davydov components at about 1.84\,eV and 1.98\,eV) to the larger excitation dipoles for acenes. However, it has been shown recently that neither the Davydov splitting nor the exciton dispersion in acene crystals can be explained on the basis of molecular excitations only, but they can only be understood when an admixture of charge transfer (CT)
excitations to the molecular (Frenkel) excitons is taken into account \cite{Yamagata2011}.

\begin{figure}
 \centering
\includegraphics[width=0.8\linewidth]{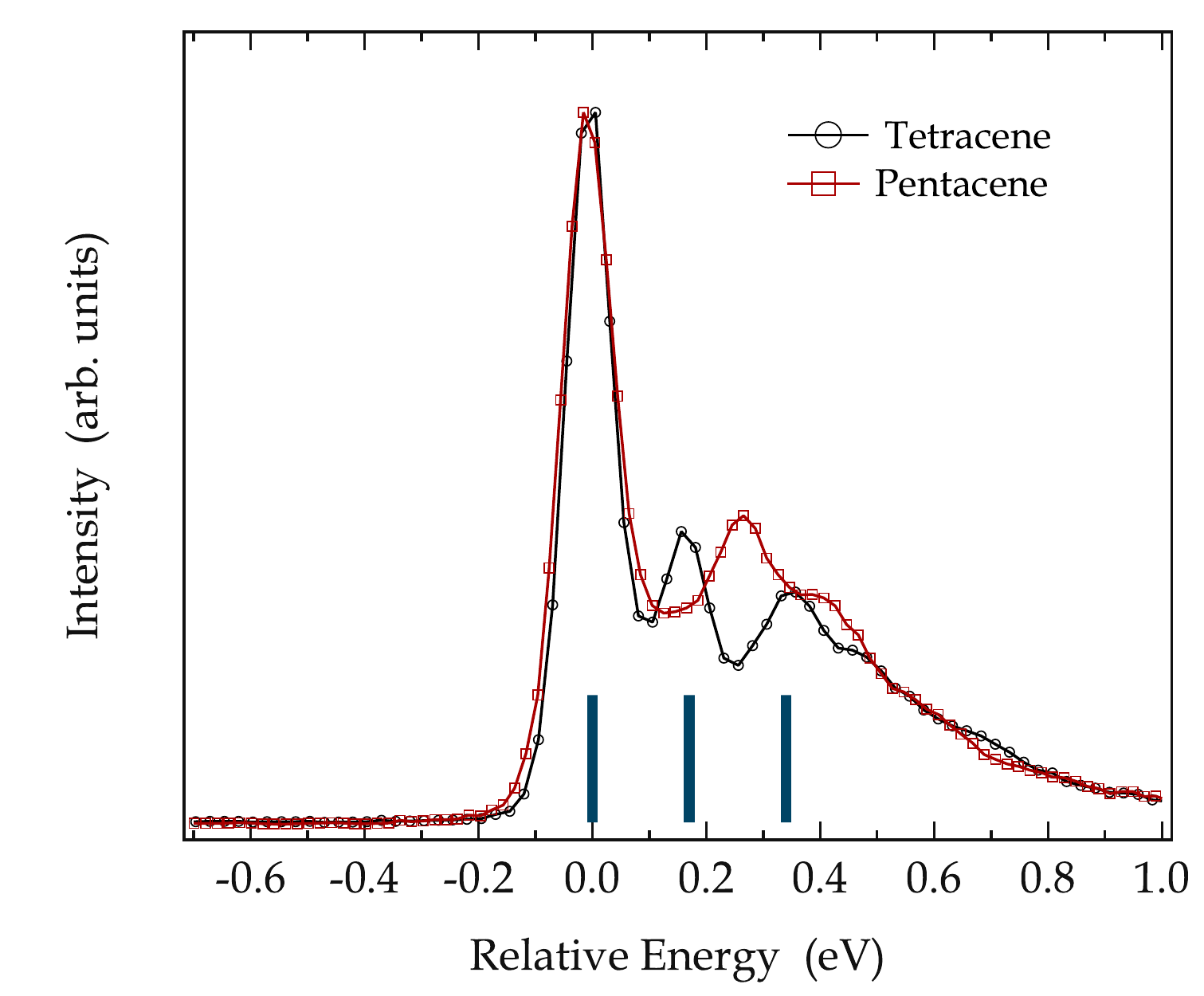}
\caption{Comparison of the exciton structure of tetracene and pentacene measured with $q$ parallel to the $a^*$-axis. The energy position of the main excitation and vibrational satellites in the optical absorption spectra of tetracene and pentacene in solution \cite{Bree1960,Zanker1969} are indicated by vertical bares.}
\label{f3}
\end{figure}

In the following we would like to concentrate on the contribution of CT excitons to the acene spectra as shown above. In Fig.\,\ref{f3} we
depict a comparison of the excitation spectra for tetracene and pentacene with the momentum vector parallel to the corresponding $a^*$ axis. These spectra have been shifted in energy such that the first excitation features coincide and they are normalized to the same peak height of the first feature. The excitation spectrum of tetracene consists of the main peak followed by equidistant vibrational satellites as has been also observed on previous optical studies of crystals and films \cite{Tanaka1965,Tavazzi2008}. Moreover, this vibrational progression is also seen in optical absorption data of tetracene in solution \cite{Bree1960}, i.\,e. for individual molecules, and the energy distance of the satellites of about 170\,meV corresponds well to carbon-carbon stretching vibrations observed using Raman scattering \cite{Carreira1986}. Thus, at least on a qualitative level the tetracene excitation spectrum can be rationalized by molecular (Frenkel) excitons that couple to molecular vibrations. The equivalent picture would now be expected for pentacene, since the corresponding vibrations as well as their impact on electronic levels should be very similar for the two molecules. Fig.\,\ref{f3} however does not support this expectation. Instead, the pentacene spectrum is characterized by a satellite feature about 270\,meV above the main excitation. This energy distance a far too large to represent a vibrational satellite, and consequently the excitation 270\,meV above the main feature must be of different origin.

\par

In the light of the discussion above and recent reports in the literature we attribute the second excitation feature in the pentacene spectrum as seen in Fig.\,\ref{f3} to CT excitations in the pentacene crystal. In general, this assignment is in agreement to electro-absorption data where a CT excitation was reported at about 2.12\,eV \cite{Sebastian1981}. Also, calculations of the electronic polarization in pentacene crystals and the energy of CT states have indicated that such CT states have a binding energy of about 0.7\,eV , which places them to almost the same energy as the molecular Frenkel excitons \cite{Tsiper2003}. Given this close excitation energy of Frenkel and CT states, they have to interact, i.e. excitation spectroscopy will probe mixed Frenkel-CT states. Indeed, recent advanced calculations of the singlet excitation spectra of acene crystals \cite{Yamagata2011} and other $\pi$-conjugated molecular crystals \cite{Hoffmann2000,gisslen2011} as well as experimental studies \cite{Knupfer2002} demonstrated that the exact spectral shape can only be understood with the inclusion of CT excitons and their coupling to the molecular Frenkel states. Upon the basis of these calculations it has further been discussed that also the Davydov splitting and the exciton dispersion significantly depend on this coupling, and a description of the experimental values only is possible with the inclusion of a substantial CT contribution \cite{Yamagata2011}. Interestingly, these calculations also predict that the contribution of the CT excitations to the lowest exciton feature quite significantly varies as a function of the length of the acene molecule. For tetracene a 27 \% contribution was reported, while for pentacene this contribution is with 48 \% much larger. Such a difference must then also affect the higher lying part of the excitation spectrum, since a mixture/hybridization of Frenkel and CT states will result in two mixed states (of bonding and anti-bonding character). The resulting mixed character of these states naturally governs the spectral weight of the corresponding excitation taking into account that (most likely) the transition dipole of molecular Frenkel excitations is much large than that of the CT transitions. Thus, in a mixed system the higher lying (former pure CT) excitation gains in intensity with increasing degree of mixture due to the Frenkel contribution in its wave function. This now harbors the explanation of the difference of the excitation spectra of tetracene and pentacene as revealed in Fig.\,\ref{f3}. The interaction in tetracene is not large enough to modify the higher lying part of the spectrum visibly as compared to pure molecular excitations (as seen in solution), while in pentacene a ``new'' excitation feature shows up at about 270\,meV
above the lowest singlet exciton, and we attribute this to the anti-bonding part of the mixed Frenkel-CT system. Since both parts of the exciton wave function couple to the vibrations of the molecules, the total spectral shape of the electronic excitation spectrum becomes a complex mixture of satellites of electronic as well as vibronic origin, and a detailed analysis requires sophisticated, state-of-the art modelling, which is beyond the scope of this contribution.

\par

It is interesting to note that calculations based upon density functional theory (DFT) have predicted that the energetically lowest singlet
excitons in pentacene have a predominant charge transfer character, in good correspondence to what we have discussed above \cite{Tiago2003,Hummer2005,Hummer2005_2,Sharifzadeh2012}. Moreover, these DFT calculations also predict a dependence of the exciton wave function and its charge-transfer character on the length of the acene molecules \cite{Hummer2005}. However, the exact spectral shape as measured has not been reproduced by such calculations.

\par

Finally, coming back to the phenacenes there is an intriguing aspect with respect to the importance of CT states in the excitation spectra.
While for chrysene all low energy features as observed in Fig.\,\ref{f2} or optical data of crystals have a corresponding counterpart in the absorption spectra of individual molecules in solution \cite{Gallivan1969}, this is not the case for picene. In the latter case the 3rd excitation feature at about 3.6\,eV is not observed in solution \cite{Gallivan1969} and in addition shows a momentum dependent intensity variation that is substantially different from all other low energy excitations \cite{Roth2011_exci}. Therefore, the excitation at about 3.6\,eV in picene has been attributed to a different origin, a predominant CT excitation \cite{Roth2011_exci}. This conclusion suggests that there is a kind of universal length dependence of the importance of CT states in the excitation spectra for both the acenes as well as the phenacenes, since only for the longer representatives, picene and pentacene, the CT excitations modify the spectra visibly.

\section{Summary}

The low energy excitation spectra of chrysene and picene as well as tetracene and pentacene crystals have been investigated using electron
energy-loss spectroscopy measurements at 20\,K. Our data reveal a significant difference between the former and the latter two materials. While for the phenacenes the excitation onset is characterized by up to five weak excitation features with only small anisotropy and without visible Davydov splitting within the $a^*$, $b^*$ planes, the acene spectra are dominated by a large excitation close to the onset and a sizable Davydov splitting. We further show that the spectral shape of the pentacene excitation spectrum provides clear evidence for a large admixture of molecular Frenkel-type excitons with charge-transfer excitations resulting in excited states with a significantly mixed character. This conclusion is in good agreement with recent advanced calculations which predicted a charge-transfer admixture to the lowest singlet excitation which is significantly dependent upon the length of the acene molecules. Moreover, also for picene and chrysene we observe differences which point towards an increased charge-transfer contribution to the singlet excitation spectrum in the former.

\begin{acknowledgments}
 We thank R. Sch\"onfelder, R. H\"ubel and S. Leger for technical assistance. This work has been supported by the Deutsche
Forschungsgemeinschaft (grant number KN393/14).
\end{acknowledgments}

\end{document}